# A Reputation System for Multi-Agent Marketplaces

Anton Kolonin, Ben Goertzel, Cassio Pennachin, Deborah Duong, Matt Iklé ,
Nejc Znidar, Marco Argentieri

SingularityNET Foundation, Amsterdam, Netherlands
{anton, ben, cassio}@singularitynet.io

## Abstract

We present an exploration of a reputation system based on explicit ratings weighted by the values of corresponding financial transactions from the perspective of its ability to grant "security" to market participants by protecting them from scam and "equity" in terms of having real qualities of the participants correctly assessed. We present a simulation modeling approach based on the selected reputation system and discuss the results of the simulation.

## 1 Introduction

The latest developments in e-Commerce and the emergence of global commercial online ecosystems with world-wide connectivity based on the Internet makes it critical to assess the reliability of the vendors and suppliers in these ecosystems in reliable way [Zheng and Jin, 2009]. Multiple design solutions for reputation systems serving the purpose exist, such as discussed by other authors: [Swamynathan *et al*., 2010], [Sänger and Pernul, 2018]. The most critical part of any system intended for this purpose appears to be the ability to handle a high degree of anonymity [Androulaki *et al*., 2008] of market participants, as it is characteristic of modern distributed [Gupta *et al*., 2003] ecosystems, including ones that are based on public networks [Blömer *et al*., 2015] such as blockchains.

In particular, in this work, we rely on a reputation system design based on the "weighted liquid rank" concept [Kolonin *et al*., 2018] and an implementation of the concept [Kolonin *et al*., 2019] applied to generic multi-agent marketplaces. Given the existing implementation of the system itself and the simplistic simulation described in the latter publication, in scope of this work we provide a more advanced simulation modeling of the same system and investigate market conditions which make it possible for honest market participants to be granted "security", protecting them from scam, and "equity", ensuring that their qualities are assessed fairly in the market.

## 2 Reputation System Implementation

Most of the "weighted liquid rank" design of the reputation system per [Kolonin *et al*., 2018] and its implementation according to [Kolonin *et al*., 2019] are used in this work, and we here we expand upon these earlier findings, that the reputation system based on explicit ratings weighted by financial values provides the best combination of "security" and "equity". In this work we focus primarily on use of only explicit ratings weighted by financial values. It should be noted that, as in the latter work, it has been found that a ratio of 10% suppliers and 90% consumers with no overlap is typical in generic online marketplaces, and because of this the reputation of consumers as raters cannot be assessed accordingly, thus the "liquid" part of the algorithm is not effectively applicable in such a case and only the "weighted" is employed.

---

**Algorithm 1** Weighted Liquid Rank (simplified version)

**Inputs**:
1) Volume of rated transactions each with financial value of the purchased product or service and rating value evaluating quality of the product/service, covering specified period of time;
2) Reputation ranks for every participant at the end of the previous time period.
**Parameters**: List of parmeters, affecting computations - default value, logarithmic ratings, conservatism, decayed value, etc.
**Outputs**: Reputation ranks for every participant at the end of the previous time period.
1: **foreach of** transactions **do**
2:     **let** *rater_value* be rank of the rater at the end of previous period of default value
3:     **let** *rating_value* be rating supplied by trasaction rater (consumer) to ratee (supplier)
4:     **let** *rating_weight* be financial value of the transaction of its logarithm, if logarithmic ratings parameter is set to true
5:     **sum** *rater_value*rating_value*rating_weight* for every ratee
6: **end foreach**

7: **do** normalization of the sum of the muliplications per ratee to range *0.0-1.0*, get *differential_ranks*
8: **do** blending of the old_ranks known at the end of previous peiod with differential_ranks based on parameter of conservatism, so that *new_ranks = (old_ranks\*conservatism+N\*(1-differential_ranks))*, using decayed value if no rating are given to ratee during the period
9: **do** normalization of *new_ranks* to range *0.0-1.0*
10: **return** *new_ranks*

We have explored two alternative open-source implementations of the reputation system based on the referenced design. The first one was based on the *Reputationer* class in Aigents Java project at https://github.com/aigents/aigents-java/ with a Python adapter for the SingularityNET reputation system prototype as in the *aigents_reputation_api.py* script of the https://github.com/singnet/reputation/ project. The second version was based on a native Python implementation as the *reputation_service_api.py* script in the latter project.

During the simulation modeling, the impact of the following parameters of the reputation system were explored:
- $R_d$ - default initial reputation rank;
- $R_c$ - decayed reputation in range to be approached by inactive agents eventually;
- *C* - conservatism as a blending "alpha" factor between the previous reputation rank recorded at the beginning of the observed period and the differential one obtained during the observation period;
- *FullNorm* – when this boolean option is set to *True* the reputation system performs a full-scale normalization of incremental ratings;
- *LogRatings* - when this boolean option is set to *True* the reputation system applies *log10(1+value)* to financial values used for weighting explicit ratings;
- *Aggregation* - when this boolean option is set to *True* the reputation system aggregates all explicit ratings between each unique combination of two agents with computes a weighted average of ratings across the observation period;
- *Downrating* - when this boolean option is set to *True* the reputation system translates original explicit rating values in range *0.0-0.25* to negative values in range *-1.0 to 0.0* and original values in range *0.25-1.0* to the interval *0.0-1.0*.
- *UpdatePeriod* – the number of days to update reputation state, considered as observation period for computing incremental reputations.

## 3 Reputation System Simulation

Based on a market analysis reported on in the referenced work [Kolonin *et al*., 2019] where it was affirmed that a ratio of 9:1 between buyers and sellers (9 times as many buyers than sellers) was the most realistic, we used this ratio as a fixed market parameter. With this ratio, we simulated populations of 1000 agents interacting over 6 months on an everyday basis. We also implemented feedback, making the reputation system available to consumers for choosing suppliers, who then rate them in the same reputation system.

During initial simulations, *80%* of the population were "good" agents with fair market behaviors and the remaining *20%* were "bad" agents that commit scams. It was also assumed that the amount of "bad" agent transactions is higher than the amount of "good" agent transactions as a result of the "bad" agents pumping up the reputation of "fake" suppliers as they pose as consumers. It was assumed that an average "bad" agent spends substantially less on every transaction, compared to an average "good" agent. Specifically, in the simulations it was assumed that "bad" agents emit *10* times more transactions a day than the good ones, but the ratio of the average transaction payment amount was varied. For the latter ratio, we considered three types of conditions: an "unhealthy" market with a ratio equal to *10* (so the "good" agents have just *10* times more costly purchases), a "semi-healthy" one with a ratio of *20* and a "healthy" one with a ratio of *100*.

The simulation, the code of which is at https://github.com/singnet/reputation/tree/master/agency/python/src/snsim/reputation, is meant to resemble a market place, in which there are different product or service categories that have a variety of prices reflective of the market. Each supplier agent has an intrinsic ability to satisfy its customers with its product or service, and ratings of each supplier are normally distributed around this "expected goodness" ($R_{cea}$ as defined further), here drawn from a normal distribution with mean of 0.75 and standard deviation of 0.15 to N(0.75, 0.0225), bucketed into five ratings of 0.0, 0.25, 0.5 0.75 and 1.0. Scammer agents are consistently given a zero by good agents. Each agent has different (normally distributed) needs for every product or service, and shops for what it needs the most first. There is a normally distributed time until each product or service is needed again. There is a limit to how much shopping an agent can do in a day, and the number of active agents can vary. Agents have differing (normally distributed) propensities to try out new agents, different tolerances for staying with the old ones, and different capacities to forget. Bad agents have a normally distributed ring size of other bad agents which they (mutually) rate positively, here drawn from N(8,4)

**Algorithm 2** Market Simulation

**Input**: Consumer and Supplier trade behaviors

**Output**: Metrics, Agent qualities, transactions, ranks

1: Assign Agents to Behaviors based on Normal Random Variates
2: Every day for 6 months:

Each consumer makes shopping list
Agents drop past suppliers according to satisfaction
　If reputation system in use:
　　If "winner take all" usage:
　　　Agents choose new suppliers with the highest
　　　　reputation score
　　Else if "roulette wheel" usage:
　　　Agents choose new suppliers in proportion to
　　　　their reputation scores
　　Else if "thresholded random" usage:
　　　Agents choose new suppliers randomly over a
　　　　reputation score threshold
　　Agents make purchases and rate suppliers
　Else if reputation system not in use:
　　If agents have no experience with suppliers
　　　Agents choose new suppliers randomly
　　Agents make purchases and rate suppliers
3: **Print** metrics

## 4 Performance Metrics

In this work we extend the range of the metrics used in referenced research [Kolonin et al., 2019]. The first key metric measures are the economic parameters of the market such as volume ratio for market spending by "good" agents versus "bad" agents ($V_g / V_b$), the percentage of funds spent by "good" agents to "bad" scammers or "loss to scam" ($C_{gbg}$), and the ratio between the earnings of scammers and their spendings or "profit from scam" ($C_{gbb}$). The latter financial metrics serve to identify the amount of financial resources that fair consumers lose in scams, or the relative amount that scammers would earn by cheating the reputation system, so changes in these values based on use of the reputation system can actually provide the business value of using it.

Further we measure utility, or how happy fair consumers are with their purchases. It is the average rating given to purchases, regardless of the kind of purchase. This metric would shift with different rules for ratings, such as a different default rating, however, within a scenario this metric has the same relative value.

$$Utility = \frac{\sum_i \sum_j F_{ij}}{N} \qquad (1)$$

where $F_{ij}$ is the sum of all ratings made by good agents and $N$ is the number of ratings made by all of the good agents, i.e. all agents which have an expected goodness of $R_{ea} = 1$, to be

$$PCCW = \frac{cov(x,y,w)}{\sqrt{cov(x,x,w)*cov(y,y,w)}} \text{ where } cov(a,b,w) = \frac{\sum w(x-avg(x,w))(y-avg(y,w))}{\sum(w)} \qquad (2)$$

discussed further on.

We also explored metrics that assess the reputation system's prediction of the quality of each agent to be either good or bad, or ability of the reputation system to have a high accuracy for both assessments, following the referenced work [Kolonin et al., 2019]. These metrics compare distributions of expected reputations $R_{ea}$ per agent $a$ against lists of computed reputations $R_{ca}$ for the same set of agents. In our work these values are measured on continuous scale, so $R_{ea}$ is used to assess the reputation system's ability to predict the level of "goodness" of an agent. The "bad" agents are expected to have $R_{ea}$ close to *0.0*, while the "good" agents are expected to have $R_{ea}$ substantially greater than *0* and up to *1.0*.

The standard Pearson correlation coefficient *PCC* between $R_{ca}$ and $R_{ea}$ is used in the referenced work [Kolonin et al., 2019]. However, it makes it possible to evaluate only overall quality of prediction so if the *PCC* value is below 1.0 then one can't tell if it is because the flawed "security" or missed "equity". Because of the reason, we have also used a "weighted Pearson correlation coefficient," weighted so that agents matter in proportion to how good they are, or how bad they are, in order to indicate the capacity of the reputation system to be used for security (by being good at determining which agents may be scammers) vs. the capacity of the reputation system to judge the quality of the honest agent's products fairly, so that they may participate proportionately in the economy.

where $avg(y,w)$ is a weighted average. For *PCCG*, the *PCC* weighted towards good, $w$ is the $(R_{cea})$ vector. For *PCCB* it is the $(1-R_{cea})$ vector. We report the weighted average of the market volume of the *PCC*, *PCCG* and *PCCB* per product or service category, called *PCC* by category, *PCCG* by category, and *PCCB* by category. This is useful because we are ultimately looking for a way to compare providers of a product or service category, and PCC is a good way to measure the ability to rank.

The standard confusion matrix metrics for recall, precision, F1 and accuracy have been used to assess the "security" as effectiveness of the reputation system for the primary reputation system goal of protecting the public from scams. That is, reputation system was considered a tool to solve "classification problem" categorizing agents as either "good" or "bad", in the sense of agent being a fair one or a scammer, respectively. Here, a positive result would mean either a "good" agent is categorized as "good" in the sense that it is not a scammer when it is scored over a threshold or a "bad" agent categorized as "bad", respectively. It is a counted as a true positive if the agent is in fact good, and is a false positive if the agent is in fact bad. On the other hand, if an agent is scored under the threshold and is in fact good it is a false negative, and if the agent scores under the threshold and is in fact bad, it is a true negative. In our work, we have used the threshold as 40% reputation rank value on scale between 0% and 100%, based on search for the optimal threshold value differentiating the "good" and the "bad" to the greater extent.

## 5 Simulation Results

We present three scenario experiments on the reputation system to find the sets of parameters in which the reputation system performs best as well as the boundary values of those parameters. The first is an experiment is a test of the parameters under the conditions of a market that is close to equilibrium, that is, a market in which there is neither a serious oversupply or serious shortages. Next, in order to test the robustness of the reputation system we look at its usage and presentation, in the usual case, where agents trade with suppliers in proportion to their ratings, and additionally extremes where agents choose the top-rated supplier, as well as where they simply choose a random supplier over a threshold. Finally, to see if the reputation is robust with respect to market volume surges we take the champion parameters of the equilibrium market and use them to test another setting of the reputation system: the periodic setting, where the update period is set longer than a day, in contexts with and without periodic surges in purchasing.

The simulation was focused on the case where the reputation system with explicit ratings weighted by financial transactions is used. For this case we have explored different combinations of parameters evaluating different metrics to access the level of "security" and "equity" provided by the reputation system.

The extent to which system can provide "security" and "equity" is illustrated by Fig.1 below. We clearly see that we can identify "bad agents" with loss of 2 "false negatives". At first glance one would think that this system only separates the scammers from the honest agents, providing "security" while not giving fair ratings to the quality of the honest agents: it seems that it cannot identify the expected goodness so the "equity" is not granted. However, the next figure shows that when you take each product or service category into account separately, it does predict an agents ratings accurately.

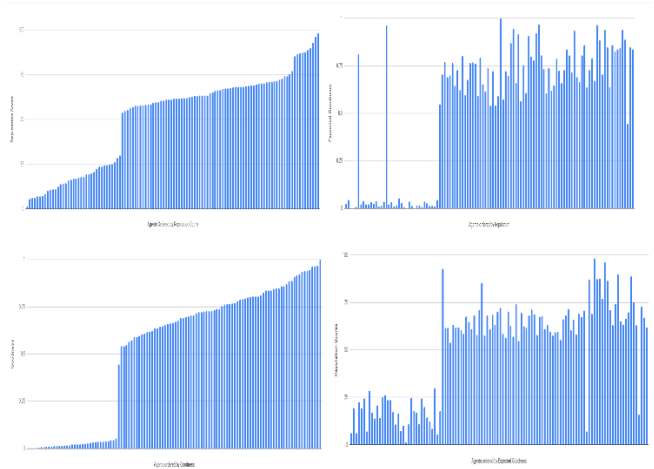

Figure 1. This is an illustration of the suppliers from a sample run of 1000 agents, 10% of which are suppliers. Each bar is a single supplier. The reputation system parameters are FullNorm=True, Weighting=True, LogRatings=False, $R_d$=0.5, C=0.5, $R_c$=0.0, Downrating=False. On the upper left we present computed goodness ($R_{ca}$) and on the upper right we present expected goodness($R_{ea}$), both ordered by computed goodness. On the lower left we present expected goodness ($R_{ea}$) and the lower right we present computed goodness ($R_{ca}$), both ordered by expected goodness. In the charts on the upper and lower right, scammers appear on the left, and honest agents on the right.

Indeed, the Fig.2 presents how the expected goodness can be evaluated on per-category basis, assuming that evaluation of reputation is domain-specific [Kolonin *et al*., 2018], according to specific price ranges and sub-markets of specific kind of goods or services of the greater market. Where there are more suppliers and thus more market volume, regardless of whether bad agents are present, the calculated goodness are more predictive of the expected goodness of the agent. The predictions are accurate wherever there are more than three samples. This result shows a consistent clear demarcation between bad agents on the left of each chart and the good agents on the right.

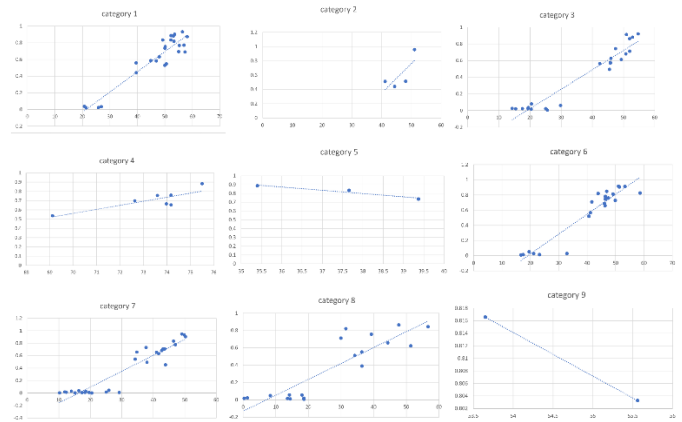

Figure 2. These charts from a single sample run show the calculated goodness (Rca) along the X axis and the expected goodness value (Rea) on the Y axis, separated by product or service category. They include the product or service categories of the realistic simulation that have more than one supplier. Rca is more predictive of Rea the larger the sample size (the slope approaching 1), and is accurate wherever the sample size is over three.

| MVR | reputation system? | conservatism | default | precision | recall | f1 | pearson by category | pearson g by category | pearson b by category | loss to scam | profit from scam | utility | utility change |
|---|---|---|---|---|---|---|---|---|---|---|---|---|---|
| unhealthy | Not Used | | | | | | | | | 0.06% | 2.2% | 0.55 | 0% |
| unhealthy | used | 0.1 | 0.1 | 0.71 | 0.44 | 0.55 | 0.10 | 0.18 | 0.04 | 1.28% | 36.5% | 0.56 | 1% |
| unhealthy | used | 0.1 | 0.5 | 0.79 | 0.48 | 0.60 | 0.71 | 0.75 | 0.70 | 0.48% | 13.6% | 0.58 | 6% |
| unhealthy | used | 0.1 | 0.9 | 0.72 | 0.46 | 0.56 | 0.42 | 0.47 | 0.39 | 0.53% | 14.8% | 0.58 | 5% |
| unhealthy | used | 0.5 | 0.1 | 0.78 | 0.44 | 0.56 | 0.33 | 0.37 | 0.34 | 1.32% | 37.1% | 0.55 | 1% |
| unhealthy | used | 0.5 | 0.5 | 0.83 | 0.56 | 0.67 | 0.01 | 0.08 | -0.06 | 0.08% | 2.2% | 0.58 | 6% |
| unhealthy | used | 0.5 | 0.9 | 0.82 | 0.57 | 0.67 | 0.82 | 0.83 | 0.81 | 0.08% | 2.2% | 0.58 | 6% |
| unhealthy | used | 0.9 | 0.1 | 0.93 | 0.48 | 0.63 | 0.77 | 0.77 | 0.77 | 0.07% | 2.1% | 0.58 | 5% |
| unhealthy | used | 0.9 | 0.5 | 0.93 | 0.49 | 0.64 | 0.78 | 0.80 | 0.77 | 0.08% | 2.1% | 0.58 | 6% |
| unhealthy | used | 0.9 | 0.9 | 0.91 | 0.51 | 0.65 | -0.08 | -0.04 | -0.12 | 0.08% | 2.1% | 0.58 | 5% |
| semi-healthy | not used | | | | | | | | | 0.06% | 4.4% | 0.55 | 0% |
| semi-healthy | used | 0.1 | 0.1 | 0.94 | 0.61 | 0.74 | 0.70 | 0.64 | 0.74 | 0.09% | 5.1% | 0.59 | 7% |
| semi-healthy | used | 0.1 | 0.5 | 0.89 | 0.59 | 0.71 | 0.61 | 0.57 | 0.66 | 0.08% | 4.6% | 0.59 | 7% |
| semi-healthy | used | 0.1 | 0.9 | 0.94 | 0.61 | 0.74 | 0.70 | 0.64 | 0.74 | 0.09% | 5.1% | 0.59 | 7% |
| semi-healthy | used | 0.5 | 0.1 | 0.92 | 0.56 | 0.69 | 0.67 | 0.63 | 0.70 | 0.14% | 7.7% | 0.59 | 7% |
| semi-healthy | used | 0.5 | 0.5 | 0.96 | 0.65 | 0.77 | 0.80 | 0.75 | 0.82 | 0.06% | 3.2% | 0.58 | 6% |
| semi-healthy | used | 0.5 | 0.9 | 0.96 | 0.70 | 0.81 | 0.81 | 0.75 | 0.84 | 0.06% | 3.4% | 0.58 | 5% |
| semi-healthy | used | 0.9 | 0.1 | 0.96 | 0.63 | 0.76 | -0.01 | -0.03 | -0.02 | 0.06% | 3.5% | 0.58 | 5% |
| semi-healthy | used | 0.9 | 0.5 | 0.96 | 0.63 | 0.76 | 0.75 | 0.69 | 0.78 | 0.07% | 4.1% | 0.58 | 5% |
| semi-healthy | used | 0.9 | 0.9 | 0.96 | 0.65 | 0.77 | 0.35 | 0.33 | 0.34 | 0.06% | 3.5% | 0.58 | 5% |
| healthy | not used | | | | | | | | | 0.06% | 22.2% | 0.55 | 0% |
| healthy | used | 0.1 | 0.1 | 1.00 | 0.82 | 0.90 | 0.70 | 0.61 | 0.76 | 0.03% | 8.9% | 0.58 | 6% |
| healthy | used | 0.1 | 0.5 | 1.00 | 0.86 | 0.93 | 0.13 | 0.13 | 0.06 | 0.03% | 8.3% | 0.58 | 5% |
| healthy | used | 0.1 | 0.9 | 1.00 | 0.82 | 0.90 | 0.70 | 0.61 | 0.76 | 0.03% | 8.9% | 0.58 | 6% |
| healthy | used | 0.5 | 0.1 | 1.00 | 0.81 | 0.90 | 0.83 | 0.75 | 0.86 | 0.03% | 9.2% | 0.58 | 6% |
| healthy | used | 0.5 | 0.5 | 1.00 | 0.97 | 0.99 | 0.95 | 0.92 | 0.96 | 0.04% | 12.1% | 0.57 | 4% |
| healthy | used | 0.5 | 0.9 | 1.00 | 0.95 | 0.97 | 0.90 | 0.84 | 0.92 | 0.05% | 13.7% | 0.57 | 4% |
| healthy | used | 0.9 | 0.1 | 1.00 | 0.92 | 0.96 | 0.83 | 0.75 | 0.87 | 0.06% | 16.1% | 0.57 | 4% |
| healthy | used | 0.9 | 0.5 | 1.00 | 0.97 | 0.99 | 0.90 | 0.84 | 0.92 | 0.06% | 17.7% | 0.57 | 4% |
| healthy | used | 0.9 | 0.9 | 1.00 | 0.92 | 0.96 | 0.71 | 0.63 | 0.76 | 0.06% | 16.5% | 0.57 | 4% |

Figure 3. The first of three market scenario types to test the parameterization and robustness of the reputation system. In this equilibrium scenario, the boundary value tested is the market volume ratio, and the parameters tested are the conservatism (conserv) and default parameters. MVR (Market Volume Ratio) states: healthy: over 200 semi-heathy: 75-200, unhealthy: below 75. The best performing combination, default 0.5/conservatism 0.5, is underlined.

| MVR | reputation system? | scenario | precision | recall | f1 | pearson by category | pearson g by category | pearson b by category | loss to scam | profit from scam | utility | utility change |
|---|---|---|---|---|---|---|---|---|---|---|---|---|
| unhealthy | not used | | | | | | | | 0.04% | 1.3% | 0.55 | 0% |
| unhealthy | used | winner take all | 0.93 | 0.23 | 0.36 | 0.55 | 0.69 | 0.45 | 0.09% | 2.8% | 0.59 | 9% |
| unhealthy | used | roulette wheel | 0.80 | 0.38 | 0.51 | 0.51 | 0.61 | 0.44 | 0.06% | 1.6% | 0.55 | 0% |
| unhealthy | used | thresholded random | 0.67 | 0.53 | 0.59 | 0.66 | 0.82 | 0.52 | 1.05% | 31.3% | 0.59 | 8% |
| semi-healthy | not used | | | | | | | | 0.04% | 2.5% | 0.55 | 0% |
| semi-healthy | used | winner take all | 0.93 | 0.26 | 0.41 | 0.58 | 0.71 | 0.38 | 0.03% | 1.6% | 0.61 | 11% |
| semi-healthy | used | roulette wheel | 0.98 | 0.46 | 0.62 | 0.69 | 0.71 | 0.63 | 0.04% | 2.3% | 0.56 | 2% |
| semi-healthy | used | thresholded random | 0.80 | 0.57 | 0.66 | 0.67 | 0.83 | 0.43 | 0.52% | 31.0% | 0.60 | 11% |
| healthy | not used | | | | | | | | 0.04% | 12.7% | 0.55 | 0% |
| healthy | used | winner take all | 1.00 | 0.32 | 0.49 | 0.73 | 0.72 | 0.71 | 0.02% | 7.2% | 0.61 | 11% |
| healthy | used | roulette Wheel | 1.00 | 0.82 | 0.90 | 0.95 | 0.90 | 0.93 | 0.03% | 9.0% | 0.55 | 1% |
| healthy | used | thresholded random | 1.00 | 0.62 | 0.76 | 0.85 | 0.86 | 0.83 | 0.04% | 10.8% | 0.61 | 12% |

Figure 4. In the second market scenario type we test the robustness of the reputation system against both normal and extreme usages, where "Roulette Wheel" indicates agent choice in proportion to the calculated reputation values, "Thresholded Random" indicates that agents choose suppliers randomly but over a threshold, and "Winner Take All" indicates that agents choose the top ranked suppliers.

Figure 3 illustrates how, in a healthy market, the reputation system is effective in reducing the loss of consumers to scam,ensuring that nearly all of the suppliers it recommends are fair market participants while at the same time rating them fairly. At test of multiple simulation runs shows that, in a healthy market scenario, there is a 99% chance that the Loss to Scam that comsumers suffer is less when the reputation system is in use than when it is not. In our consideration of the importance of balance between offerring protection from scammers and making sure no honest suppliers are rated fairly, we have underlined the "champion" parameter combination that meets both goals, default 0.5 and conservatism 0.5.

The healthy market champion's precision of 1.0 in Figure 3 shows that the reputation system did not recommend any of the scammers despite their best efforts at pumping their ratings, while at the same time rating the non-scammers fairly as evidenced by the Pearson-good by category score of 0.92. The Pearson by category scores show that the system ranks all agents effectively within a category of goods and services. If default is set low, then we can make the consumers slightly happier with their purchases as a result of the reduced loss to scam, but this is at the expense of ranking good agents fairly.

Figure 4 illustrates the effectiveness of the reputation system under different usages. One possible usage is winner take all, when reputation system scores are presented to consumers such that they would probably just take the highest scoring supplier. Such a usage has increased the happiness that consumers have in thier products (increased utility by 11%), however more studies need to be done in other scenarios that have enterers and leavers, as winner take all dyanamics have been known to close markets to newcomers and thus ultimately result in lower utility. The healthy market with winner take all dynamics was poor in ranking fair suppliers (Pearson-good by category of 0.71). However, our champion scenario, which models the case where customers are presented with the reputation system such that they would select agents in proportion to their reputation scores, ranked non-scamming suppliers fairly (Pearson-good by category of 0.9) so they can participate in the economy. Proportionate selection usage is roulette wheel selection [Bäck Thomas, 1996], which does better than the thresholded random usage which does not rank quite as well.

| MVR | reputation system? | periodic? | period | precision | recall | f1 | pearson by category | pearson g by category | pearson b by category | loss to scam | profit from scam | utility | utility change |
|---|---|---|---|---|---|---|---|---|---|---|---|---|---|
| unhealthy | not used | yes | | | | | | | | 0.07% | 4.5% | 0.57 | 0.0% |
| unhealthy | used | yes | 1 | 0.67 | 0.69 | 0.68 | 0.77 | 0.85 | 0.66 | 0.36% | 14.6% | 0.56 | -1.0% |
| unhealthy | used | yes | 7 | 0.83 | 0.44 | 0.58 | 0.92 | 0.93 | 0.91 | 0.05% | 2.1% | 0.58 | 1.0% |
| unhealthy | used | no | 7 | 1.00 | 0.20 | 0.33 | 0.70 | 0.79 | 0.63 | 0.06% | 1.8% | 0.56 | -1.0% |
| semi-healthy | not used | yes | | | | | | | | 0.07% | 9.1% | 0.57 | 0.0% |
| semi-healthy | used | yes | 1 | 0.87 | 0.71 | 0.78 | 0.83 | 0.85 | 0.77 | 0.35% | 28.7% | 0.57 | 1.0% |
| semi-healthy | used | yes | 7 | 1.00 | 0.67 | 0.80 | 0.96 | 0.95 | 0.96 | 0.04% | 3.5% | 0.58 | 1.0% |
| semi-healthy | used | no | 7 | 1.00 | 0.28 | 0.44 | 0.84 | 0.83 | 0.80 | 0.06% | 3.2% | 0.56 | -1.0% |
| healthy | not used | yes | | | | | | | | 0.07% | 45.3% | 0.57 | 0.0% |
| healthy | used | yes | 1 | 1.00 | 0.82 | 0.90 | 0.94 | 0.89 | 0.94 | 0.06% | 25.1% | 0.56 | -1.0% |
| healthy | used | yes | 7 | 1.00 | 0.94 | 0.97 | 0.98 | 0.95 | 0.99 | 0.04% | 15.8% | 0.58 | 1.0% |
| healthy | used | no | 7 | 1.00 | 0.78 | 0.87 | 0.97 | 0.89 | 0.97 | 0.04% | 11.6% | 0.56 | -1.0% |

Figure 5. In the third market scenario type, the period that the reputation system is updated is tested, in scenarios that are periodic (the default) in that there is a surge of purchasing every seven days for half of the goods, and scenarios which are. Scenarios with no reputation system are presented for comparison in all scenario types, and best performing parameter combinations in both healthy and semi-healthy markets, is underlined.

Figure 5 illustrates how the reputation system performs well even in markets that have periodic surges. The champion of the market is the champion of the entire paper, where the reputation system which updates only every week , on a market with weekend purchasing surges, has a 0.98 Pearson-good by category score. In markets without surges, the same system performs nearly as well, and still reduces loss to scam over the reputation system without the update period and over no reputation system.

Based on studies within the scope of performed simulations, the following observations were made.

- The best set of reputation parameters are: *Weighting=True, LogRatings=False, $R_d$=0.5, C=0.5, $R_c$=0.0, Downrating=False.*
- Parameters that matter to the result are *Weighting, $R_d$, C, and $R_c$. FullNorm* does not matter to the result, and *Downrating and LogRatings* consistently detract from the result.
- The reputation system cuts the loss to scam (LTS) that consumers suffer nearly in half. In a healthy market LTS is cut from 0.06% to 0.03 or 0.04%.
- The reputation system gives an excellent ability to rank good agents by product or service category and discern bad agents in a healthy market, with Pearson-good by category scores over 0.9 and precision of discerning bad agents of 1.0, and little societal loss to scam (< 0.0003). The capability of the reputation is retained in the semi-healthy market, but decreases dramatically in the unhealthy market (Pearson of 0.61) making the semi-healthy market the borderline condition for the reputation system to function.
- The reputation system can handle periodic sales (such as seasonal and weekly surges in purchasing) with the same ability to rank good agents and discern bad (Pearson scores >= 0.9 and precision of 1.0). If there is no periodicity in purchases, but the periodic mode of the reputation system is on, the performance is comparable (Pearson scores = 0.89 and precision of 1.0), however, if there is a periodicity in purchases and periodic mode is not employed, performance is slightly lower (Pearson scores = 0.83 and precision of 1.0).
- The reputation system performs best when used by agents to trade with suppliers in proportion to their calculated goodness scores (Precision 1.0, Pearson scores >=0.9) but still works adequately at the extremes of "Winner Take All" choice of suppliers (Precision 1.0, Pearson scores >= 70) and random above threshold choice (Precision 1.0, Pearson scores >= 80)
- It may be possible to incentivize agents to trade with suppliers in proportion to their calculated goodness scores financially, not only to keep the reputation system accurate, but in the interest of equity.
- Further studies need to be done on automatically finding a threshold under which an agent is never recommended. Dynamic adjustment of the threshold will make the system even more robust against market conditions.

# 6 Conclusion

Based on the presented work, we can make the following conclusions.

- For the studied market conditions, the use of the reputation system based on explicit ratings weighted by financial values of transactions provides both "security" in the form of protection from scam and "equity" for the honest participants of the market
- The parameters that we have found may be used for fine-tuning of the real reputation systems and the metrics that we define may be used to optimize these parameters in further studies.
- The options suggested for threshold management, and for the incentivization of users to choose amongst the listed suppliers in proportion to their reputation scores may be considered as an extra application layer on top of the reputation system.

The following issues will be considered for our further work.

- Make the simulations more realistic, exploring larger scales of markets above 1000 agents and 6 months, studying wider ranges of possible market conditions, with different populations of scamming agents, different ratios of transactions committed by the agents and implementing different temporal patterns of the behaviors of the agents, primarily having different rates of newcomers and leavers.
- Evaluate possibility to improve "security" and "equity" with use of rating aggregation.
- Explore possibility of implementation of dynamic reputation threshold management enabling to cut off "most likely scam" from "most likely honest" and study its effect on "security" and "equity".
- Explore possibility of implementation and evaluation of the incentivization of the honest agents to act as members of "reputation militia" so the consumers are choosing suppliers in proportion to their scores.
- Implementation and simulation of the staking-based reputations [Kolonin *et al.*, 2018] and exploration of its impact on improvement of "equity", especially in the case with newcomers entering the system and leavers leaving the one at realistic rates.